\documentclass{article}
\usepackage{spconf,amsmath,graphicx}

\usepackage{booktabs}
\usepackage{float}

\def\ps@IEEEtitlepagestyle{%
	\def\@oddfoot{}%
	\def\@evenfoot{\mycopyrightnotice}%
}
\def\mycopyrightnotice{%
	{\footnotesize This paper is accepted to be published in: 2018 IEEE International Conference on Image Processing, Oct 7-10, 2018, Athens, Greece.
	\par\ \ \ \footnotesize IEEE Copyright Notice: \copyright IEEE 2018 Personal use of this material is permitted. Permission from IEEE must be obtained for all other uses, in any current or future media, including reprinting/republishing this material for advertising or promotional purposes, creating new collective works, for resale or redistribution to servers or lists, or reuse of any copyrighted component of this work in other works.
\hfill}
	\gdef\mycopyrightnotice{}
}
\newcommand\blfootnote[1]{%
	\begingroup
	\renewcommand\thefootnote{}\footnote{#1}%
	\addtocounter{footnote}{-1}%
	\endgroup
}

\title{OSLO: Automatic Cell Counting and Segmentation for Oligodendrocyte progenitor cells}
\name{Haoyi Ma$^1$, Rebecca Beiter$^2$, Alban Gaultier$^2$, Scott T. Acton$^1$ and Zongli Lin$^1$
\thanks{This work was supported in part by the US Army Research Office under grant W911NF1510275.}
}

\address{$^1$Charles L. Brown Department Of Electrical and Computer Engineering, University of Virginia\\$^2$Department of Neuroscience, University of Virginia\\Charlottesville, VA, USA}
\begin{document}
%
\maketitle
\begin{abstract}
Reliable cell counting and segmentation of oligodendrocyte progenitor cells (OPCs) are critical image analysis steps that could potentially unlock mysteries regarding OPC function during pathology. We propose a saliency-based method to detect OPCs and use a marker-controlled watershed algorithm to segment the OPCs. This method first implements frequency-tuned saliency detection on separate channels to obtain regions of cell candidates. Final detection results and internal markers can be computed by combining information from separate saliency maps. An optimal saliency level for OPCs (OSLO) is highlighted in this work. Here, watershed segmentation is performed efficiently with effective internal markers. Experiments show that our method outperforms existing methods in terms of accuracy.
\end{abstract}
\begin{keywords}
saliency detection, cell counting, bio-image analysis  
\end{keywords}
\blfootnote{\mycopyrightnotice}
\vspace{-0.6cm}
\section{Introduction}
\label{sec:intro}
Oligodendrocyte progenitor cells (OPCs) are a specific class of glia cells which is most commonly known as the resident pool of progenitors \cite{dawson2003ng2}. Recent evidence suggests that OPCs may play an integral role in disorders such as depression \cite{wang2014white}. OPC count and morphology enable the understanding of how OPCs respond to pathological conditions (i.e., proliferation vs death vs differentiation). The traditional approach to OPC detection and counting relies mainly on experienced experts to mark and count OPCs manually, which is extremely tedious and time consuming. 

Graylevel thresholding is a common approach used to discriminate foreground from the background, which can be attempted in cell counting tasks given simple, limited situations \cite{embleton2003automated}\cite{davis2004real}\cite{bi2015semi}. However, thresholding approaches are sensitive to noise and lack efficacy in complex scenarios in which the graylevel histogram is not bimodal \cite{acton2009biomedical}. For OPCs, a more complex approach that exploits the salient structure of the cells is required.

\begin{figure}[htb]
\centering
\includegraphics[width=88mm,height=55mm]{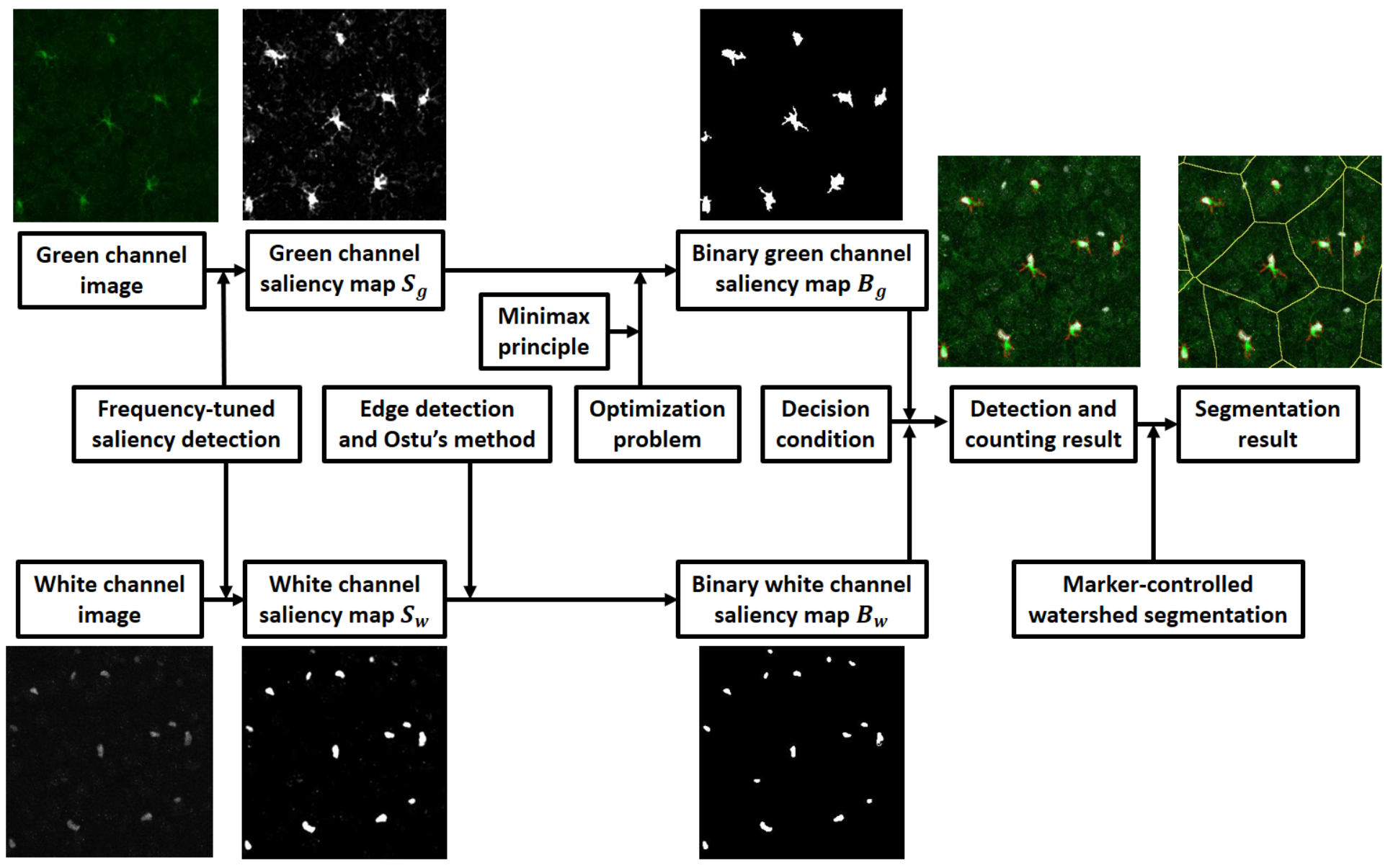}
\caption{An overview of the OSLO approach.}
\label{Fig1}
\end{figure}

For unsupervised machine learning methods, counting problems are tackled by performing grouping based on self-similarities \cite{ahuja2007extracting} or motion similarities \cite{rabaud2006counting}. However, the accuracy of such fully unsupervised methods is limited. In terms of supervised learning formulation, Lempitsky \textit{$et \ al$}. \cite{lempitsky2010learning} provided a supervised learning framework that focuses on the practically-attractive case in which the training images are manually annotated. However, providing sufficient annotations for the training data is still an extremely tedious and time consuming task.

Image saliency detection is a common method to detect and segment significant objects under the complex background \cite{srivatsa2015salient}\cite{mukherjee2015saliency}. Saliency detection can be used for cell counting and segmentation because microscopic images of cells are usually captured by highlighting the cells as foreground objects. Zheng \textit{$et \ al$}. \cite{zheng2017robust} performed cell counting and segmentation for microscopic images of non-setae phytoplankton species using saliency detection. Pan \textit{et \ al}. \cite{pan2012leukocyte} developed a framework for segmentation of leukocytes using a novel saliency detection method. One main issue related to saliency-based methods is that the accuracy of counting and segmentation heavily relies on the saliency level used in binarizing the saliency map \cite{luo2010saliency}, which remains a difficult problem.

In this paper, we propose a saliency-based method to count and segment OPCs automatically. As shown in Fig. 1, first, saliency maps are computed on separate white and green channels (to be described in Section 3) with a frequency-tuned saliency detection algorithm. By combining information from two separate saliency maps specific to the OPC detection problem, we obtain the final detection results and internal markers. The saliency level selection of the green channel saliency map is formulated as an optimization problem with linear weight constraint, the weight parameter is computed using the minimax principle and without training (optimal saliency level for OPCs or OSLO). With the internal markers, segmentation can be performed efficiently with the marker-controlled watershed algorithm.

The remainder of this paper is organized as follows. Section 2 introduces priors for the frequency-tuned saliency detection algorithm. Section 3 elaborates on the implementation of our proposed algorithm. Section 4 compares our algorithm with other algorithms and verifies the accuracy of our method. Finally, contributions of this paper and future work are discussed in Section 5.
\vspace{-0.1cm}
\section{Frequency-tuned saliency detection}
\label{sec:format}
Achanta \textit{$et \ al$}. \cite{achanta2009frequency} proposed a frequency-tuned (FT) saliency detection algorithm. The basic idea is to filter image from low frequency to high frequency using several band-pass filters \cite{runxin2015survey}. The final full resolution saliency map is computed by combining the outputs of all these band-pass filters. 

The whole process is realized by composing several difference of Gaussian (DoG) filters, a summation over several narrow band-pass DoG filters with standard deviations in the ratio $\rho$ results in:
\vspace{-2mm}
\begin{eqnarray}
\sum_{n=0}^{N-1}G(x,y,{\rho}^{n+1}\sigma)-G(x,y,{\rho}^n\sigma)&&\nonumber\\
=G(x,y,{\rho}^N\sigma)-G(x,y,\sigma)&&
\end{eqnarray}
for an integer $N\ge 0$, which is simply the difference of two Gaussians for which the ratio of standard deviations is equal to ${\rho}^N$.

The first Gaussian standard deviation ${\rho}^N$ is driven to extremely big to implement a large ratio in standard deviations and results in a notch in frequency at DC while preserving all other frequencies. A small Gaussian kernel is used to remove high frequency noise and textures \cite{achanta2009frequency}.

The saliency map $S$ for an image $I$ can be formulated as:
\vspace{-2mm}
\begin{equation}
    S(x,y) = \|I_{\mu} - I_{\omega_{hc}(x,y)}\|
\end{equation}
where $I_{\mu}$ is the mean image feature vector, $I_{\omega_{hc}(x,y)}$ is the corresponding image pixel vector value in the Gaussian blurred version of the original image, and $\|\cdot\|$ is the $L_2$ norm.  

FT can compute a full resolution saliency map while keeping the calculation speed fast. The resulting saliency map has uniformly highlighting salient regions and clear boundaries. Also, FT can be used for the detection of relatively small objects.
\vspace{-0.2cm}
\section{Automatic cell detection and segmentation}
\label{sec:pagestyle}
\vspace{-0.1cm}
\subsection{Image characteristics}
In order to selectively label OPCs, platelet-derived growth factor receptor A (PDGFRA) is used in conjunction with a transcription factor Olig2. PDGFRA marks the cell surface of OPCs while Olig2 marks the nuclei of OPCs. After fluorescent labeling of PDGFRA and Olig2 proteins, z-stacks that permit the quantification of cellular composition and OPC morphology can be obtained by using lasers of varying wavelengths to excite different fluorescent proteins. Individual images of PDGFRA and Olig2 are pseudo-colored as green and white and compiled into a combined channel image.

An OPC has many branches, called \textit{processes}, that extend out from the cell body. While OPCs generally maintain their own territory \cite{hughes2013oligodendrocyte}, the processes are connected when analyzing a z-stack of images projected into 2 dimensions. As shown in Fig. 2, the low contrast of nuclei and connected processes make it difficult to mark the OPCs in the combined channel image. Therefore, it is crucial to incorporate information from separate channels to complete the counting and segmentation tasks. 
\begin{figure}[htb]
\begin{minipage}[b]{0.3\linewidth}
  \centering
  \centerline{\includegraphics[width=28mm,height= 24mm]{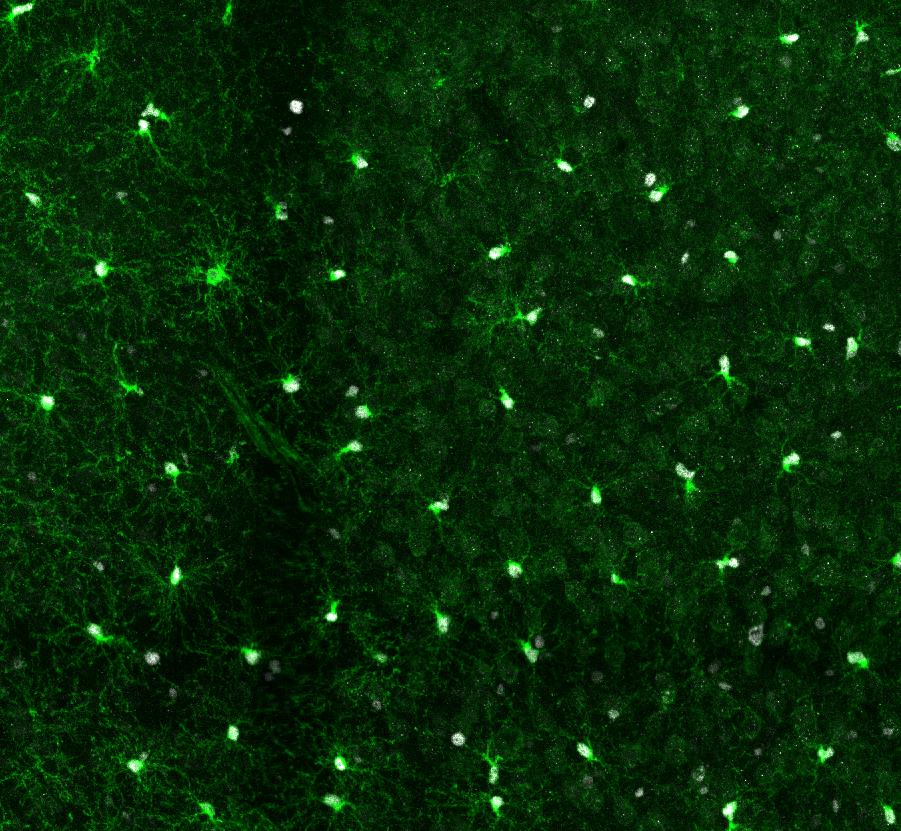}}
  \centerline{(a)}\medskip
\end{minipage}
\hfill
\begin{minipage}[b]{.3\linewidth}
  \centering
  \centerline{\includegraphics[width=28mm,height= 24mm]{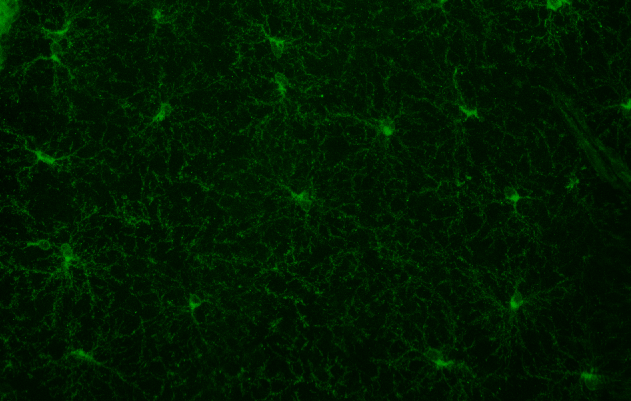}}
  \centerline{(b)}\medskip
\end{minipage}
\hfill
\begin{minipage}[b]{0.3\linewidth}
  \centering
  \centerline{\includegraphics[width=28mm,height= 24mm]{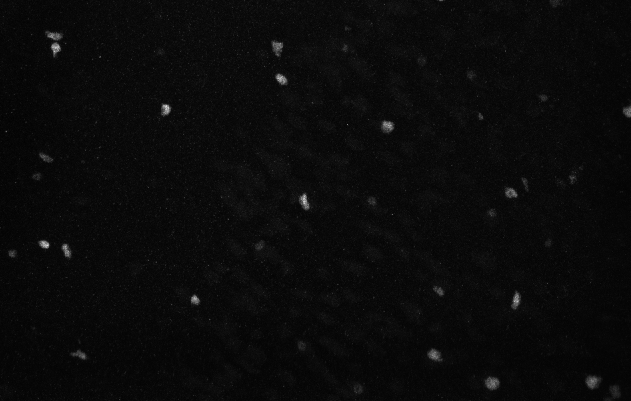}}
  \centerline{(c)}\medskip
\end{minipage}
\vspace{-0.3cm}
\caption{(a) Combined channel image. (b) Green channel image. (c) White channel image.}
\label{fig:res}
\end{figure}
\vspace{-0.6cm}
\subsection{Salient region detection}
\subsubsection{Saliency detection in white channel}
For the white channel image, we apply FT to suppress the noise in the background and get the saliency map $S_w(x,y)$ of white channel, where $(x,y)$ correspond to the spatial coordinate of original white channel image. To obtain a more accurate detection and estimation of the area of the nuclei, we combine Canny edge detection and a binary map computed using Ostu's threshold $T_{ostu}$. 
The binary map $B_{w1}$ can be obtained using $T_{ostu}$ on the saliency map $S_w$. With Canny edge detection, we can compute the edge map of $S_w$ as $E_w$. By filling all the holes of $E_w$, another binary map $B_{w2}$ can be obtained. The final binary map of the white channel saliency map is computed as 
\vspace{-2mm}
\begin{equation}
    B_w(x,y) = B_{w1}(x,y)|B_{w2}(x,y).
\end{equation}

\vspace{-0.6cm}
\subsubsection{Saliency level selection and saliency detection in green channel}
The intersecting regions between binary maps of the green channel and the white channel saliency maps are the primary cue for OPC detection and counting. The saliency level should be set sufficiently low such that details of the cell body are retained, which makes the ratio between the area of the nucleus and the area of the corresponding intersecting region smaller. However, if the saliency level is set too low, processes of different OPCs will be connected, and the ratio between the area size of cell body and the area size of corresponding nucleus will be relatively large. 

To find an appropriate saliency level of the green channel saliency map, we formulate the saliency level selection problem as an optimization problem:
\vspace{-2mm}
\begin{equation}
    L_g\! = \!\underset{L}{\arg\min}\ \lambda \frac{\sum\limits_{m=1}^{M}{R_{wi|L}(m)}}{M}+(1-\lambda) \frac{\sum\limits_{m=1}^{M}{R_{gw|L}(m)}}{M},\!\!
\end{equation}
where $M$ is the number of intersecting regions between the binary map $B_g$ of the green channel saliency map using saliency level $L$ and the binary map $B_w$ of the white channel saliency map, $R_{wi|L}(m)$ is the ratio between the area size of the nucleus in the white channel and that of the intersecting region correspond to the $m$th intersecting region when the saliency level of the green channel saliency map is $L$, and $R_{gw|L}(m)$ is the ratio between the area size of the cell body in the green channel and that of the intersecting region correspond to the $m$th intersecting region when the saliency level of the green channel saliency map is $L$.
\vspace{-0.2cm}
\subsubsection{Parameter selection using minimax principle}
With the saliency level selection problem formulated as an optimization problem, the remaining task is to choose $\lambda$ in the optimization equation. We apply the minimax principle to avoid the problem of extremely low or high weights \cite{gennert1988determining}. To explain our parameter selection method, we rewrite our cost function as 
\vspace{-2mm}
\begin{equation}
    E(\lambda, L) = \lambda R_1(L) + (1-\lambda) R_2(L),\ 0\le\lambda\le1,
\end{equation}
where $L$ is the saliency level of the green channel saliency map. For a given value of $\lambda$, there exists a value of $L$ which minimizes $E(\lambda,L)$. Let the minimum value be 

\vspace{-2mm}
\begin{equation}
    E^{*}(\lambda) = \underset{L}{min}\ {E(\lambda,L)} = E(\lambda,L^{*}(\lambda)).
\end{equation}

Given $\lambda$, $L^{*}(\lambda)$ is computed as the value with the minimum total cost. The main difficulty in choosing $\lambda$ is that if it is either too low or too high, one of the objective functions will be inadequately represented in the total cost. We use the minimax principle to minimize the possible loss for a worst case scenario. That is, we find $\lambda^{*}$ such that $E^{*} = E^{*}(\lambda^{*})$ is maximized,

\vspace{-2mm}
\begin{equation}
    \lambda^{*} = \underset{\lambda}{\arg\max}\ E^{*}(\lambda).
\end{equation}

Using $\lambda^{*}$ as the parameter, we can compute $L_g$ as the saliency level of the green channel saliency map by solving the optimization problem we defined in equation (4). The binary map $B_g$ of the green channel saliency map is computed using $L_g$ as shown in Fig. 3 (a).  
\begin{figure}[htb]
\begin{minipage}[b]{0.31\linewidth}
  \centering
  \centerline{\includegraphics[width=28mm, height = 24mm]{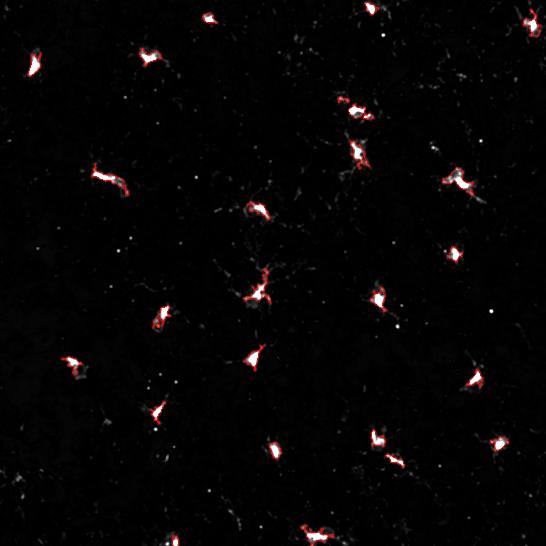}}
  \centerline{(a)}\medskip
\end{minipage}
\hfill
\begin{minipage}[b]{.31\linewidth}
  \centering
  \centerline{\includegraphics[width=28mm, height = 24mm]{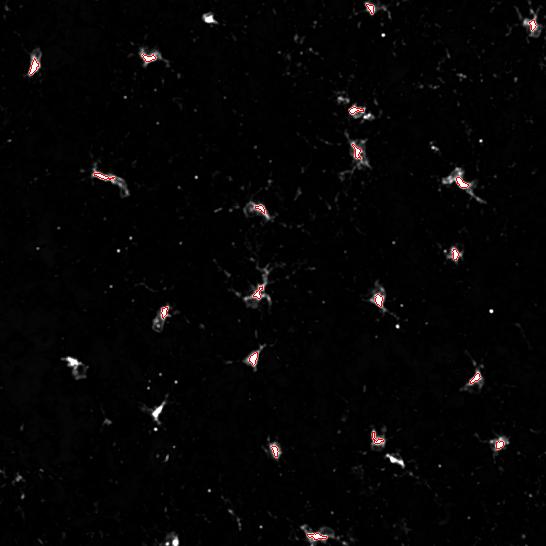}}
  \centerline{(b)}\medskip
\end{minipage}
\hfill
\begin{minipage}[b]{0.31\linewidth}
  \centering
  \centerline{\includegraphics[width=28mm, height = 24mm]{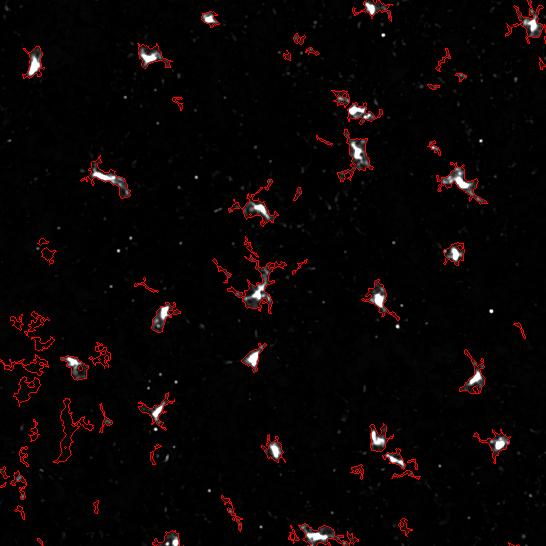}}
  \centerline{(c)}\medskip
\end{minipage}
\caption{Binary maps of the green channel saliency map using (a) our saliency level selection method, (b) Ostu's threshold selection method and (c) Bradley's threshold selection method.}
\label{fig:res}
\end{figure}
\vspace{-0.7cm}
\subsubsection{Cell counting and segmentation}
With the binary maps $B_w$ and $B_g$ we get from separate white and green channels, we compute the binary map $B_c$ of cell candidates as:
\vspace{-2mm}
\begin{equation}
    B_c(x,y) = B_w(x,y)|B_g(x,y).
\end{equation}

The upper bound of the ratio between the area size of a nucleus and the area size of the corresponding intersecting region is computed as:
\vspace{-2mm}
\begin{equation}
R_{uwi|L_g} \!\!= \!\!\bar{R}_{wi|L_g} \!\!+ \!3\sqrt{\frac{1}{M}\!\sum\limits_{m=1}^{M}(R_{wi|L_g}(m)\!\! -\!\!\bar{R}_{wi|L_g})^2},
\end{equation}
where $\bar{R}_{wi|L_g}$ is the mean value of ${R}_{wi|L_g}$. 

If $R_{wi|L_g}(m)\le R_{uwi|L_g}$, we count the $m$th union region as one OPC. Using all the united regions that satisfy the requirement as internal markers, a marker-controlled watershed algorithm is applied for the final cell segmentation.

\section{EXPERIMENTS}
\label{sec:typestyle}
\vspace{-0.1cm}
\subsection{Experimental settings}
Using the minimax principle, we calculate the value of $\lambda$ and compute the saliency level of the green channel saliency map. By using the marker-controlled watershed algorithm, the final segmentation result can be obtained. We refer to our method as optimal saliency level for OPCs (OSLO).
\vspace{-0.1cm}
\subsection{Quantitative results}
We evaluate our algorithm by counting the number of OPCs within 15 intravital microscopic images of our dataset hand labeled by experts and compute $F_1$ score as $F_1 = 2\cdot\frac{precision\cdot recall}{precision+recall}$. The $F_1$ score is within $[0,1]$, a larger $F_1$ score means a more accurate detection and counting result. Table 1 shows the results of our algorithm compared with those by other methods.
\begin{center}
\textbf{Table 1}\ \ $F_1$ score of different methods.\\
\begin{tabular}{cccccc} \toprule
No.  &  Ostu        &  Canny     & Bradley         & OSLO \\ \hline
\#1  & 0.73       & 0.57     & 0.91          & 0.97      \\
\#2  & 0.33       & 0.43     & 0.93          & 0.98       \\
\#3  & 0.30       & 0.35     & 0.94          & 0.98        \\
\#4  & 0.57       & 0.76     & 0.87          & 0.96         \\
\#5  & 0.41       & 0.44     & 0.93          & 0.98          \\
\#6  & 0.49       & 0.54     & 0.96          & 0.99           \\
\#7  & 0.25       & 0.46     & 0.90          & 0.97           \\
\#8  & 0.39       & 0.53     & 0.86          & 0.98           \\
\#9  & 0.52       & 0.42     & 0.91          & 0.98            \\
\#10  & 0.33      & 0.47     & 0.90          & 0.97            \\
\#11  & 0.42      & 0.55     & 0.81          & 0.98              \\
\#12  & 0.42      & 0.48     & 0.93          & 0.98          \\
\#13  & 0.35      & 0.50     & 0.85          & 0.96             \\
\#14  & 0.48      & 0.58     & 0.87          & 0.97            \\
\#15  & 0.38      & 0.58     & 0.78          & 0.96    \\
\textbf{Ave}    & 0.42 & 0.51  & 0.89        & $\mathbf{0.97}$          \\
\bottomrule
\end{tabular}
\end{center}

Ostu's threshold selection method \cite{otsu1979threshold} and Canny edge detection are implemented using the default parameters given by Matlab. Bradley corresponds to the local adaptive threshold selection method proposed in \cite{bradley2007adaptive}. As shown in Table 1, our proposed algorithm OSLO yields the best performance compared to other methods in terms of the $F_1$ score.
\vspace{-0.1cm}
\subsection{Qualitative results}
As shown in Fig. 4 and Fig. 5, we can accurately detect and segment the OPCs even when cell bodies are extremely close and processes are strongly connected, which shows the robustness of the OSLO compared to other methods.
\begin{figure}[htb]
\centering
\includegraphics[width=85mm,height=34mm]{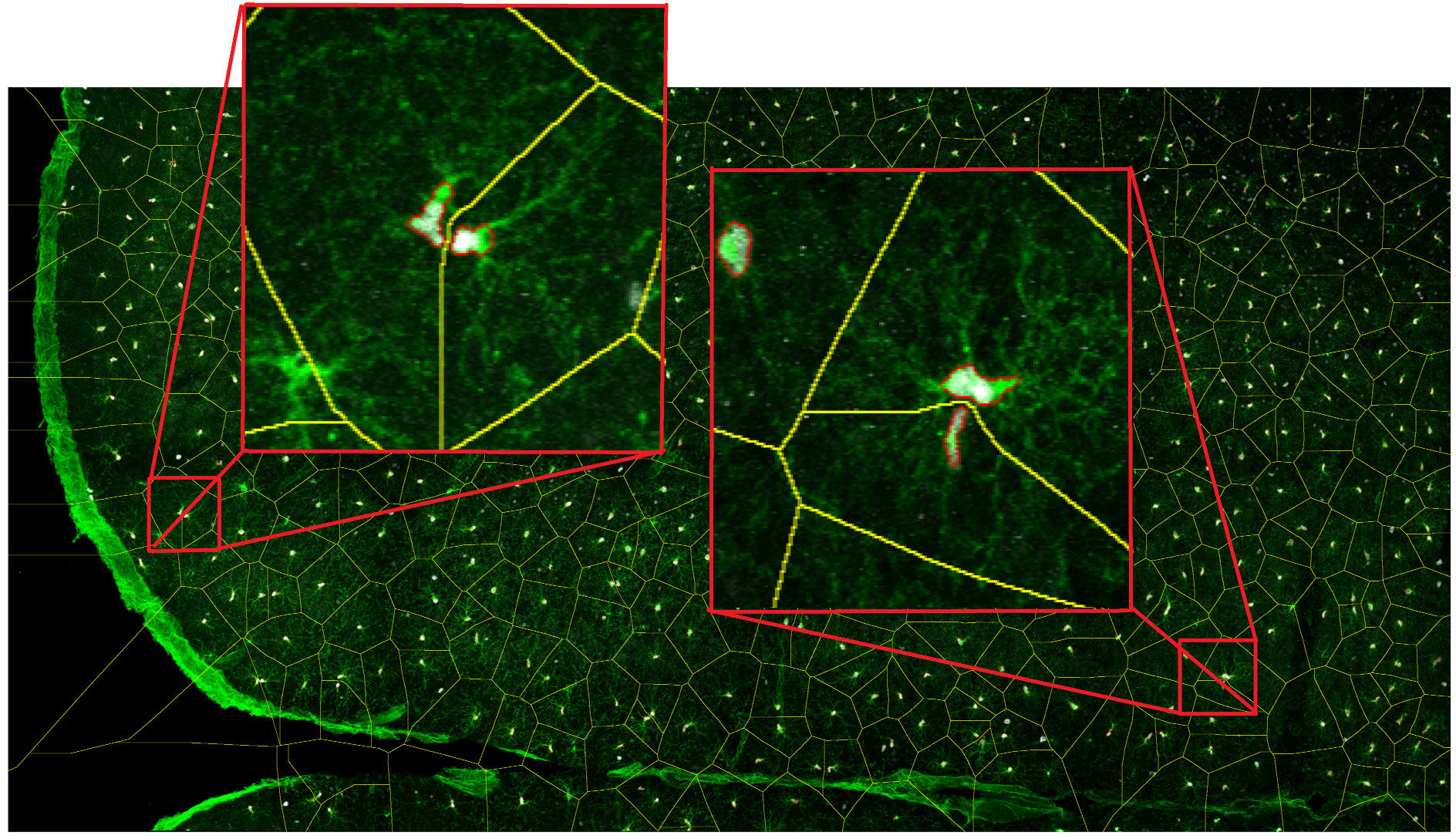}
\caption{Detection and segmentation results using OSLO and zoomed-in results of extremely close cell bodies with connected processes.}
\label{Fig1}
\end{figure}

\begin{figure}[htb]
\begin{minipage}[b]{0.31\linewidth}
  \centering
  \centerline{\includegraphics[height=26mm,width=28mm]{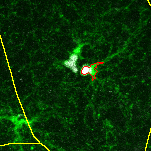}}
\end{minipage}
\hfill
\begin{minipage}[b]{.31\linewidth}
  \centering
  \centerline{\includegraphics[height=26mm,width=28mm]{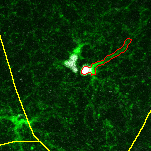}}
\end{minipage}
\hfill
\begin{minipage}[b]{0.31\linewidth}
  \centering
  \centerline{\includegraphics[height=26mm,width=28mm]{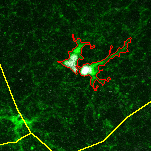}}
\end{minipage}
\begin{minipage}[b]{0.31\linewidth}
  \centering
  \centerline{\includegraphics[height=26mm,width=28mm]{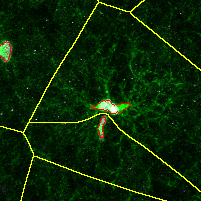}}
  \centerline{(a)}\medskip
\end{minipage}
\hfill
\begin{minipage}[b]{.31\linewidth}
  \centering
  \centerline{\includegraphics[height=26mm,width=28mm]{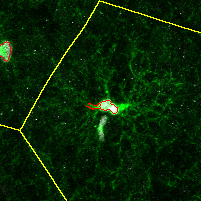}}
  \centerline{(b)}\medskip
\end{minipage}
\hfill
\begin{minipage}[b]{0.31\linewidth}
  \centering
  \centerline{\includegraphics[height=26mm,width=28mm]{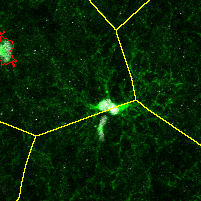}}
  \centerline{(c)}\medskip
\end{minipage}
\caption{Detection and segmentation results using (a) Ostu's threshold selection method, (b) Canny edge detection and (c) Bradley's threshold selection method correspond to the zoomed-in regions of Fig. 4.}
\label{fig:res}
\end{figure}

We can see in Fig. 6 that a cell with low contrast in the white channel cannot be classified as an OPC or not in the combined channel image even by the trained technician. However, OSLO obtains the correct result as it utilizes information from multiple channels.
\begin{figure}[htb]
\centering
\includegraphics[width=85mm,height=38mm]{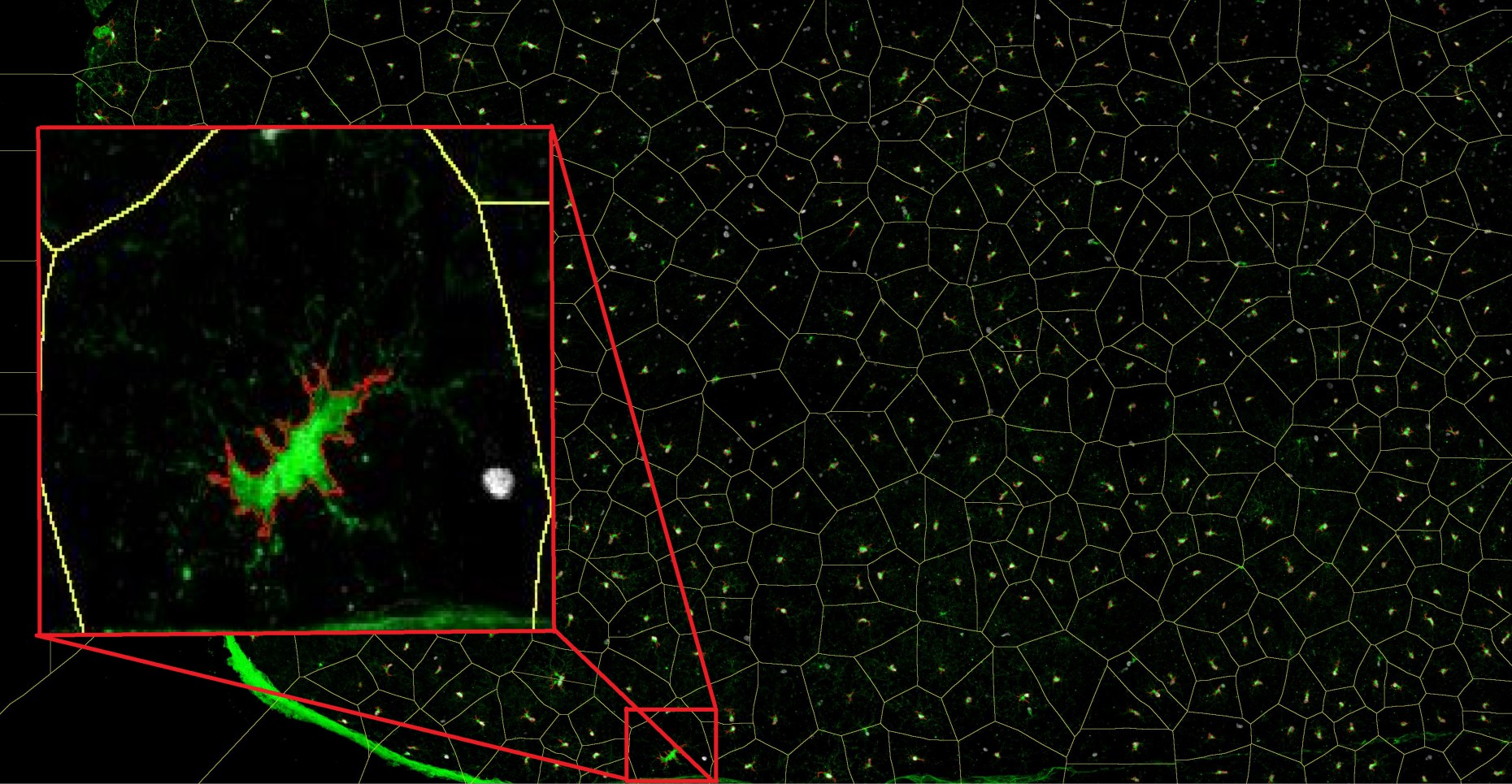}
\caption{Detection and segmentation results and zoomed-in version of a cell with a nucleus that exhibits low contrast.}
\label{fig:res}
\end{figure}

\section{CONCLUSIONS}
We proposed a novel saliency-based method to count and segment OPCs automatically. The contributions of this paper are threefold. First, a saliency-based cell counting and segmentation framework is proposed. Second, the saliency level selection problem for the green channel saliency map is formulated and solved as an optimization problem. Third, the weight parameter is computed using the minimax principle without training.

Future work includes extending the OSLO to time sequences for the enhancement of cell tracking algorithms such as \cite{goobic2005image}, \cite{ray2002active} and \cite{cui2006monte}.


\bibliographystyle{IEEEbib}
\bibliography{main}

\end{document}